\documentclass[a4paper, 11pt]{article}
\usepackage{graphicx}
\usepackage{lineno}
\usepackage{siunitx}
\usepackage{hyperref}
\usepackage{amsmath,amssymb}
\usepackage{bm}
\usepackage{authblk}

\newif\ifarxiv
\arxivtrue

\newenvironment{methods}{%
    \section*{Methods}%
    \setlength{\parskip}{11pt}%
    }{}

\newenvironment{addendum}{%
    \setlength{\parindent}{0in}%
    \small%
    \begin{list}{Acknowledgements}{%
        \setlength{\leftmargin}{0in}%
        \setlength{\listparindent}{0in}%
        \setlength{\labelsep}{0em}%
        \setlength{\labelwidth}{0in}%
        \setlength{\itemsep}{12pt}%
        }
    }
    {\end{list}\normalsize}

\begin{document}
\title{Acoustic Macroscopic Rigid Body Levitation by Responsive Boundary Hologram}
\author[1]{Seki Inoue \footnote{inoue@hapis.k.u-tokyo.ac.jp}}
\author[1]{Shinichi Mogami}
\author[1]{Tomohiro Ichiyama}
\author[2]{Akihito Noda}
\author[3]{Yasutoshi Makino}
\author[3]{Hiroyuki Shinoda}

\affil[1]{Graduate School of Information Science and Technology, The University of Tokyo, Tokyo, Japan}
\affil[2]{Department of Mechatronics, Nanzan University, Nagoya, Japan}
\affil[3]{Graduate School of Frontier Sciences, The University of Tokyo, Chiba, Japan}

\maketitle
\begin{abstract}

Propagated acoustic waves, which generate radiation pressure, exert a non-contact force on a remote object. By suitably designing the wave field, remote tweezers are produced that stably levitate particles in the air without any mechanical contact forces\cite{Whymark:1975aa, Brandt:2001aa, Foresti_2013}. Recent works have revealed that holographic traps can levitate particles even with a single-sided wave source\cite{Lee:2009aa, Shih_Tsung_Kang_2010, Zhang:2014aa, Marzo:2015aa}. However, the levitatable objects in the previous studies were limited to particles smaller than the wavelength, or flat parts placed near a rigid wall\cite{Andrade_2016}. Here, we achieve a stable levitation of a macroscopic rigid body by a holographic design of acoustic field without any dynamic control. The levitator models the acoustic radiation force and torque applied to a rigid body by discretising the body's surface, as well as the acoustic wave sources, and optimizes the acoustic field on the body surface to achieve the Lyapunov stability so that the field can properly respond to the fluctuation of the body position and rotation. In an experiment, a 40 kHz (8.5 mm wavelength) ultrasonic phased array levitated a polystyrene sphere and a regular octahedron with a size of $\sim$50 mm located 200 mm away from acoustic elements in the air. This method not only expands the variety of levitatable objects but also contributes to microscopic contexts, such as in-vivo micromachines, since shorter-wavelength ultrasound than the size of target objects can be used to achieve higher controllability and stability.
\end{abstract}
\vspace{5mm}
Acoustic radiation pressure is a static pressure applied to an object in a propagating or standing sound wave, and can apply a force to a remote object. Acoustic tweezers trap particles at points in space by utilizing acoustic radiation pressure.  Compared with optical radiation force tweezers for nano-particles or magnetic levitation targeting ferromagnetic materials, acoustic levitation by ultrasonic waves can levitate various materials, opening up the possibility of a wide range of applications, such as crystallography\cite{Ohsaka1990UndercoolingDrops}, chemistry\cite{Santesson2004AirborneAnalysis}, microbiology\cite{Ding2012On-chipWaves} and human--computer interfaces\cite{Ochiai:2014aa, Sahoo:2016aa}.

However, previous midair acoustic levitation was limited to particles smaller than the wavelength of the sonic wave, where the wave field is designed under an assumption that sound field disturbance by the target object is negligible. An exceptional example of macroscopic object levitation is squeeze-film levitation. An ultrasonic vibrator can reduce the friction and lift an object larger than the wavelength using squeeze-film effect. However, the levitation distance is much smaller than the wavelength in typical designs\cite{UEHA200026}. Another approach is to create a standing wave between wave sources and the target object \cite{Zhao2011, Andrade_2016}. Although it has been theoretically predicted that this approach can levitate an object at a distance of half the wavelength or integer multiples thereof, no experiments have been reported that demonstrate stable levitation in a three-dimensional space beyond distances equal to the wavelength. Basically, this approach cannot stabilize the transverse orientation of the object to the wave direction.

A typical levitator for particles uses a standing waves generated by a pair of transducers or a set of a transducers and a reflector\cite{Whymark:1975aa, Brandt:2001aa, Foresti_2013}. A recent research has expanded its degrees of freedom of manipulation into three-dimensions\cite{Ochiai:2014aa}. In addition, single-sided traps formed by travelling waves have been realized recently, by extending the technique used in optical tweezers, first for lateral trapping\cite{Lee:2009aa} and also for full dimensional trapping\cite{Marzo:2015aa}.

The main approach used in acoustic tweezers is to apply an approximation model, such as the Gor'kov potential, that holds on a sphere significantly smaller than the wavelength, assuming the field is not affected by the scatterer\cite{Brandt:2001aa, Shih_Tsung_Kang_2010, Foresti_2013, Marzo:2015aa, Melde:2016aa, Marzo2017RealizationDelay-lines}. This approach is often highly effective for sparse particles because it becomes a free space problem, which is analytically solvable. Another approach is to employ a cloaking technique, where the wave avoids an object and is scattered less\cite{Zhang:2014aa}. However, it yields less force, and it has not been reported that this technique gives a force sufficient to lift a solid in the air because the acoustic radiation pressure results from the scattered energy density\cite{Westervelt:1951}. 

Levitation by shorter wavelengths than the object size will not only expand the variety of levitatable targets but will also enable precise control of small particles. Levitation of a rigid body with a large clearance while stabilizing its orientation  will be of benefit in a wide range of applications, including industrial assembly such as electronics, drug delivery, micro-machine control and multiple bodies control, and will make it possible to manipulate the object along any path, even paths that would be impossible to follow by mechanical arms due to interference.

In this paper we present a novel approach: responsive boundary holography that designs wave field on impedance boundaries that can provide desirable restoring force and torque. The formulation accurately models the radiation pressure for a macro-scale rigid body by discretising its surface and optimizes the wavefield on the surface to lift and stabilize the rigid body. The condition for achieving stable rigid body levitation is that both the force and torque are in mechanical equilibrium and, in addition, adequate restoring force and torque are generated against the positional and rotational fluctuation of the rigid body. Our levitator seeks an optimum acoustic field that satisfies this condition for an arbitrarily given phased array and rigid body. As examples, we demonstrate that a sphere with a diameter of 30 mm and a regular octahedron with an edge length of 35.4 mm (a diagonal length of 50 mm) can be stably levitated by single- and double-sided phased arrays, respectively without any dynamic feedback controls.

\begin{figure}
\centering
\includegraphics[width=\textwidth]{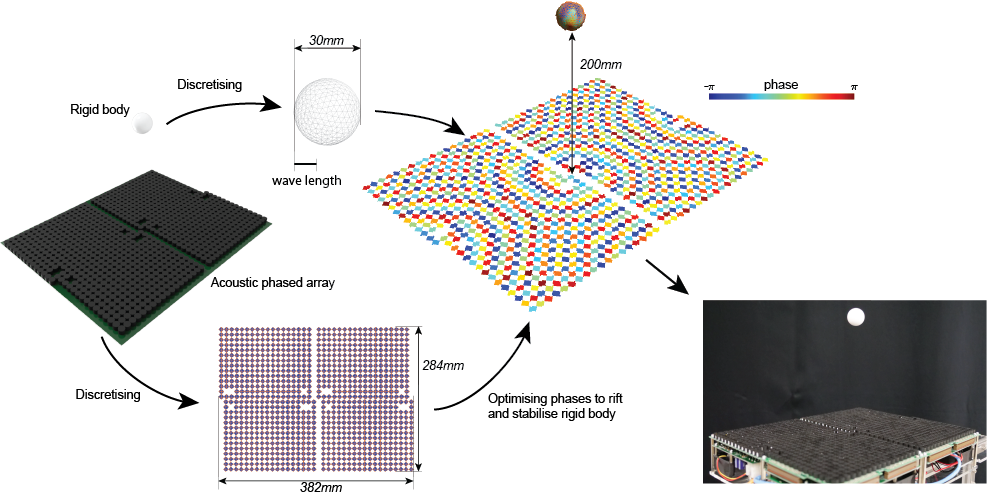}
\caption{Schematic diagram of procedure for levitating macroscopic rigid body. Geometries of acoustic phased array and target rigid body are defined. The rigid body and phased array are discretized and described as groups of passive and active acoustic elements, respectively. The phases of active elements are optimized to maximize the restoring force and torque acting on the body while giving the exact force needed to lift the body.}
\label{fig:schematic}
\end{figure}

In our approach, we derive the radiation pressure on a scatterer by discretising not only the wave sources but also the surface of the rigid body. According to the Helmholtz--Kirchhoff equation, surface elements on the scatterer are considered as passive acoustic elements. This principle is historically employed in the Boundary Element Method (BEM)\cite{Sauter2011BoundaryMethods}. As shown in Figure \ref{fig:schematic}, we describe the system with active acoustic elements (transducers) and passive acoustic elements (surface of the rigid body). Since the transducers and the rigid body are far enough apart, re-reflection on the transducers is neglected. In this formulation, the radiation force, torque and their Jacobians at the levitation point are calculated as surface integrals of the Lagrange density and are represented by a quadratic matrix form of phased array gains, which can be efficiently handled on modern parallel processors. 

The conditions for rigid body levitation are: 1) the external forces, which are typically gravity and acoustic radiation force, are balanced; 2) a restoring force is generated against positional fluctuation; 3) the net moment is zero; and 4) a restoring torque is generated against rotational fluctuation. In previous studies on particles, condition 1) is automatically satisfied by a position shift of the particle, and 3) and 4) are out of consideration. Marzo el al. solved condition 2) by maximizing the Gor'kov Laplacian to make it positive for particles\cite{Marzo:2015aa}. However, the Gor'kov potential becomes imprecise as the target becomes larger and aspherical, and the positive Laplacian is not a sufficient condition for stabilizing a rigid body even if it is small. We formulate an optimization problem to synthesize a phased array output as eigenvalue minimization with constraints. It is solvable by gradient methods such as the BFGS method. In other words, we optimize holographic surface along the boundary of the rigid body. When ignoring drag force, the rigid body levitates if and only if all eigenvalues of its Jacobian of generalized force in Method are non-positive real numbers in the neighbourhood where the linear approximation holds, which realizes the Lyapunov stability condition known in control theory (see Method, equation (\ref{eq:lyapunov})). We obtain the optimum solution by minimizing the eigenvalues keeping the force-torque equilibrium condition. It is obvious the existence of the solution is not always guaranteed and it fails to find a solution satisfying the constraints when the target is too heavy or has a too complex shape for the given phased array. In the formulation, we assume that the body boundary is sound-hard since it is valid for most rigid materials in gaseous media. The viscosity of the air is neglected in this report, though it contributes to the further stability.

\begin{figure}
\centering
\includegraphics[width=\textwidth]{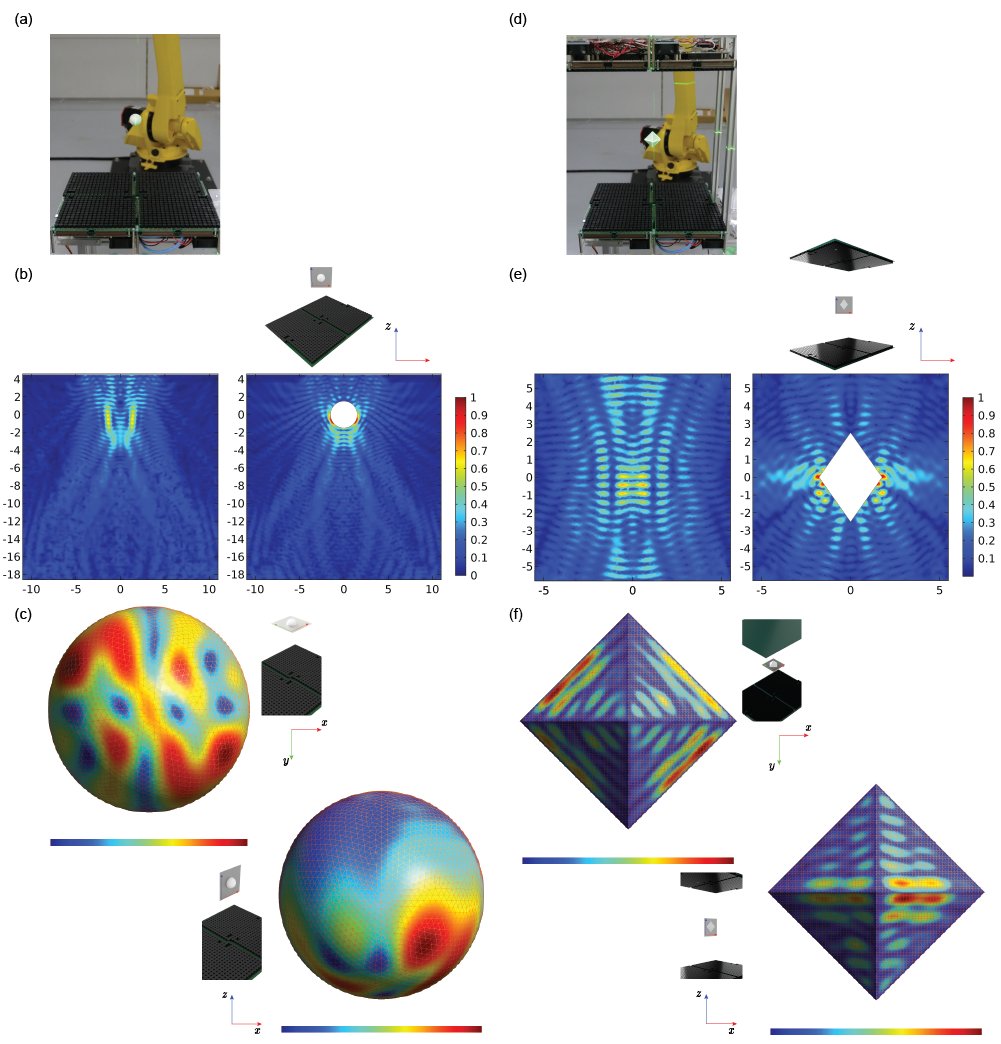}
\caption{
Levitation of a sphere and regular octahedron by 40 kHz ultrasound in air. (a) Photograph showing levitation of a sphere with a diameter of 30 mm at a point 200 mm above a single-sided phased array, which consists of 996 transducers. (b) Absolute acoustic pressure around the sphere, simulated by the finite element method on plane $x+y=0$. Wave field without the sphere (left) and wave field with the sphere (right) are shown. Note that both fields originate form an identical sound source. The pressure is normalized by the peak value of the two. (c) Absolute acoustic pressure on the surface of the sphere calculated by proposed model seen from the bottom along z-axis (left) and side along y-axis (right) (d) Photograph showing levitation of regular octahedron with diagonal length of 50 mm at a position in the middle of a double-sided phased array at a distance of 400 mm, which consists of 1992 transducers in total. (e) Absolute acoustic pressure around the octahedron simulated by finite element method on plane $x+y=0$. Wave field without the octahedron (left) and wave field with the octahedron (right) are shown. Note that both fields originate form an identical sound source. The pressure is normalized by the peak value of the two. (f) Absolute acoustic pressure on the surface of the octahedron calculated by proposed model seen from the bottom along z-axis (left) and side along y-axis (right).}
\label{fig:levit}
\end{figure}

Figure \ref{fig:levit} (a) and (d) show levitation of a sphere with a diameter of 30 mm (mass 0.6 g) and a regular octahedron with a diagonal length of 50 mm (mass 0.5 g) by 40 kHz (wavelength 8.5 mm) ultrasound in the air at a position 200 mm above the bottom phased array. 
In the figure, three green laser lines show an orthogonal basis crossing at the levitation point.
Figure \ref{fig:levit} (b) and (e) show FEM simulation of the absolute acoustic pressure on a plane perpendicular to the phased array.
The left side of each shows the acoustic field in the absence of a target rigid body, and the right side shows the one in the presence of the target rigid body. The acoustic fields are significantly changed by the rigid bodies so as to support the mass. It will be difficult for readers to recognize the acoustic fields of the left-side (without body) and right-side (with body) are originating form an identical sound source, especially in case of the octahedron. Figure \ref{fig:levit} (c) and (f) show absolute acoustic pressure on the rigid body surface calculated by our model. We can see that the pressure spreads over the surface. 

The sphere was levitated by both a travelling wave type (only by the bottom phased array, which consists of 996 transducers) and a stationary wave type (by the bottom and top phased arrays, which consist of 1992 transducers), and the latter was more stable. In the case of the octahedron, the optimizer did not reach a stable solution for the single-sided phased array while it did reach a stable solution for the double-sided phased array. In actuality, the double-sided array levitated the octahedron but the single-sided array did not in our experiments.

\begin{figure}
\centering
\includegraphics[width=\textwidth]{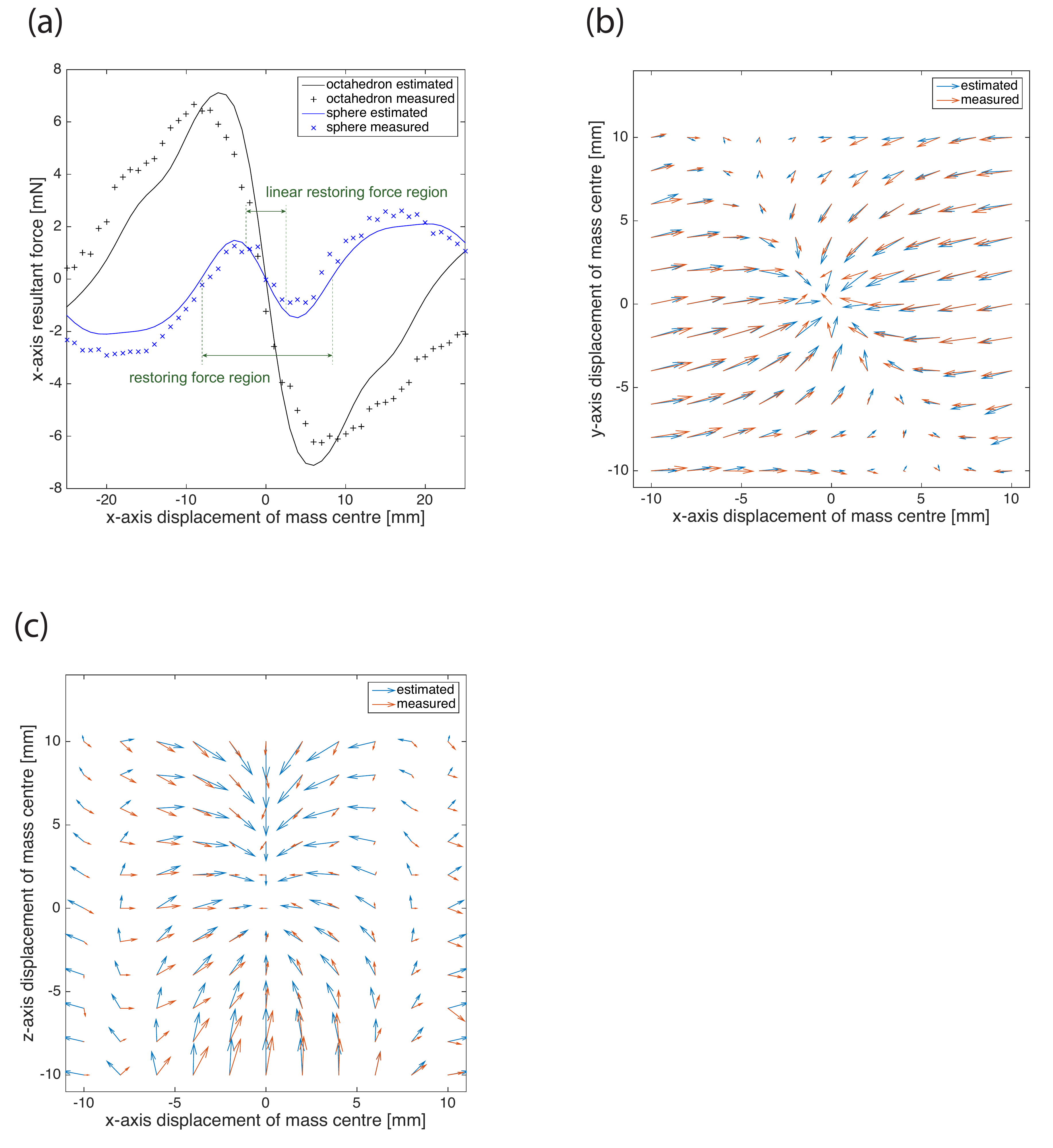}
\caption{
Estimated and measured forces applied to the rigid bodies. Configurations of rigid bodies, phased arrays and their gains are the same as in Figure \ref{fig:levit}. (a) x-axis displacement versus x-axis force applied to the octahedron (black) and the sphere (blue). Green dotted line shows the restoring force region. (b) Force vectors applied to the octahedron on horizontal plane $z=0$. Blue arrows show the estimated values by our model, and red arrows show the measured values. (c) Force vectors applied to the sphere in vertical plane $y=0$. Blue arrows show the values estimated by our model, and red arrows show the measured values.}
\label{fig:pert}
\end{figure}

Figure \ref{fig:pert} shows the estimated and measured restoring force against the displacement around the equilibrium point, the origin, for the acoustic fields and rigid objects shown in Figure \ref{fig:levit}. Figure \ref{fig:pert} (a) and (b) show the horizontal forces for horizontal displacement applied to the sphere and the octahedron. The restoring force properly acted in a direction opposite to the displacement and converged at the origin. Figure \ref{fig:pert} (c) shows the net resultant force including the restoring force and the gravity applied to the sphere in a vertical plane including the origin. The force converged at the origin in this case too, and the gradient of the force was observed along the vertical line. 

\begin{figure}
\centering
\includegraphics[width=\textwidth]{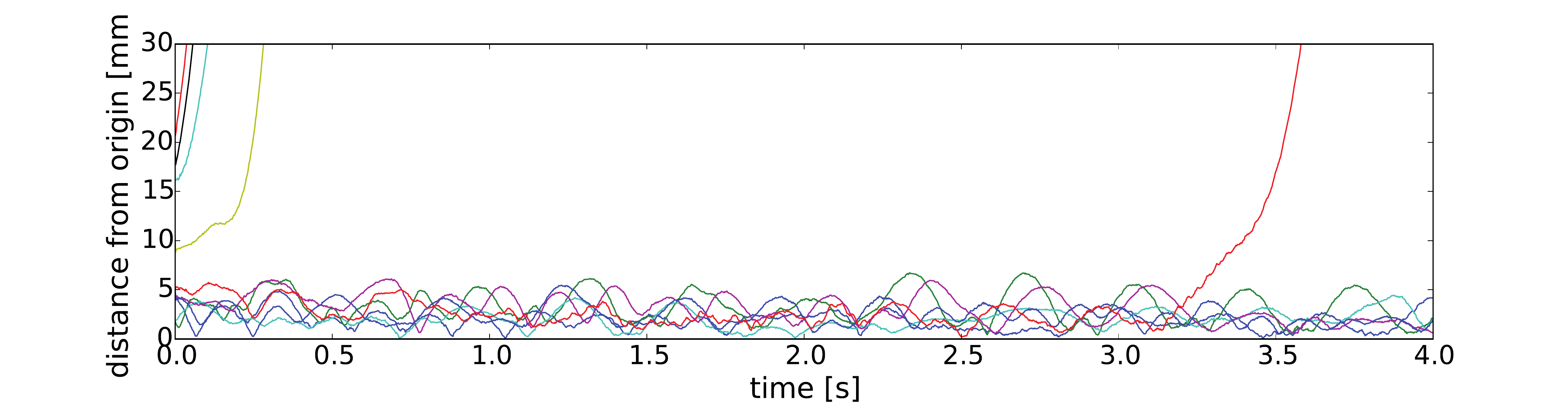}
\caption{Tracks of levitated sphere. 
996 transducers generated levitation point at distance of 200 mm for a sphere with a diameter of 30 mm. The sphere was placed at different initial positions in ten successive trials. The five spheres that dropped within 4 seconds were at initial positions more distant than 5 mm.}
\label{fig:track}
\end{figure}

The stable region of the rigid body is the neighbourhood of the levitation point where restoring force acts like linear spring since our optimizer constrains the gradient of the restoring force at the equilibrium point (see Figure \ref{fig:pert} (a)). To see this, we tracked the position of a sphere with a diameter of 30 mm from various initial points. Figure \ref{fig:track} shows the results of ten successive trials of the levitation. The different initial positions of the sphere were set as shown in the figure at $t=0$. The five spheres that dropped within four seconds were initially at positions more distant than 5 mm from the origin. The stable region of radius 5mm corresponded to the linear region in Figure \ref{fig:pert} (a). The other five spheres were levitated for 50 seconds as the minimum and 144 seconds at the maximum. Unfortunately, perfect asymptotic stability has not been achieved yet. The reason of this instability is that a lossy factor is not included in the system. In the previous microscopic particle levitation, the lossy factor is provided by the fluid viscosity. But in a macroscopic scenario with a large Reynolds number, the lack of the loss makes the system unstable.

In summary, we achieved acoustic levitation for macroscopic rigid bodies. We employed a BEM formulation for acoustic radiation pressure applied to an arbitrary smooth surface and formulated an optimization problem that maximizes the stability while balancing all forces and torques, at a point distant enough from the acoustic elements. This model can be applied to arbitrary shaped rigid bodies and arbitrarily arranged phased arrays and even metamaterials\cite{Melde:2016aa, Memoli:2017aa, Marzo2017RealizationDelay-lines}. This levitation technique is applicable to three-dimensional manipulation of multiple objects to follow complex 3D paths in the air or liquid, which would be impossible to achieve by mechanical arms. Another aspect of the presented method is it enables use of higher frequency waves for micro particle manipulation, which improves the temporal response and strengthens the holding force by steeper pressure gradient. This boundary-holographic-optimization approach is also applicable to electromagnetic fields, which improves optical tweezers\cite{Grier2003AManipulation, Shvedov:2014aa}.      

\clearpage

\clearpage
\newcommand{\dq}{\mathfrak{q}}
\newcommand{\FT}{\mathfrak{F}}
\newcommand{\pd}[2]{\frac{\partial#1}{\partial#2}}
\newcommand{\dif}[3][]{\frac{d^#1 #2}{d#3^#1}}

\begin{methods}
\subsection*{Acoustic radiation pressure on sound-hard boundary}
The force $\bm{F}$ that acts on a rigid body surface RB due to acoustic radiation pressure is expressed by a Lagrangian density $L$ and a momentum density tensor $\rho_0\bm{uu}$:
\begin{align}
\bm{F}=-\int_{\mbox{RB}} \langle L \rangle d\bm{S} + \rho_0\int_{\mbox{RB}} d\bm{S}\cdot\langle \bm{uu}\rangle,
\end{align}
\noindent where $\rho_0$ is the density of the medium, $\bm u$ is the particle velocity, and $\langle \cdot\rangle$ denotes the time domain average\cite{Westervelt:1951}.
By using the relative surface admittance $\beta$ and assuming a harmonic time-domain term, we obtain the following relation among particle velocity $\bm{u}$, sound pressure $p$ and unit surface normal $\bm{n}$:
\begin{align}
\bm{u} = -\frac{1}{jc_0k\rho_0} \nabla p, \;\bm{n}\cdot \bm{u} = -\rho_0 \beta \frac{\partial p}{\partial t} = -jk\beta p,
\end{align}
\noindent where $j$ denotes the imaginary unit, $c_0$ is the speed of sound in the medium, and $k$ is the wavenumber of the ultrasonic wave. 
We can rewrite the radiation force acting on the rigid body as:
\begin{align}
\bm{F}&=\int_{\mbox{RB}}\left( \frac{1}{2\rho_0c_0^2} \langle p^2 \rangle - \frac{1}{2}\rho_0\langle \|\bm{u}\|^2 \rangle\right) d\bm{S} + \rho_0 \int_{\mbox{RB}} \langle \bm u(d\bm{S}\cdot \bm u)\rangle \\
&=\frac{1}{2\rho_0c_0^2}\int_{\mbox{RB}}\left( \langle p^2 \rangle \bm{n} - \frac{1}{k^2}\langle \|\nabla p\|^2 \rangle \bm{n} + 2\rho_0 c_0 \beta \langle p\nabla p \rangle\right) dS.
\end{align}
Especially on a sound-hard boundary such that $\beta = 0$, this is simply:
\begin{align}
\bm{F}&=\frac{1}{2\rho_0c_0^2}\int_{\mbox{RB}} \left[ \langle p^2 \rangle -  \frac{1}{k^2}\langle \|\nabla_{\!\scriptscriptstyle\parallel} p\|^2 \rangle \right] d\bm{S},\label{eq:force}
\end{align}
\noindent where $\nabla_{\!\scriptscriptstyle\parallel}p$ means the tangential derivative at the surface. Similarly, the moment is:
\begin{align}
\bm{T}&=\frac{1}{2\rho_0c_0^2}\int_{\mbox{RB}} \left[\langle p^2 \rangle -  \frac{1}{k^2}\langle \|\nabla_{\!\scriptscriptstyle\parallel} p\|^2 \rangle \right] (\bm r - \bm c)\times d\bm{S},\label{eq:torque}
\end{align}
\noindent where $\bm r$ is the position of surface elements and $\bm c$ is the centre of mass of the rigid body.
\subsection*{Scattered acoustic wave field exerted by phased array transducers}

The pressure of an incident wave at a point $\bm r$ is described as an integral of wavefields generated by phased array elements:
\begin{align}
p_{\tiny\mbox{inc}}(\bm r)=\int_{\mbox{PAT}} q(\bm s)g(\bm r,\bm s)dS,
\end{align} 
\noindent where $g$ is Green's Function for the Helmholtz Equation, $q(\bm s)$ is the normal particle velocity on the phased array surface, and PAT denotes the surface of the phased array transducers. The Kirchhoff--Helmholtz integral theory states that the scattered wave from a rigid body surface RB is written in the following form:
\begin{align}
p_{\tiny\mbox{sca}}(\bm r) = -\int_{\mbox{RB}} \left(p(\bm s)\dfrac{\partial g(\bm r,\bm s)}{\partial n}-g(\bm r,\bm s)\dfrac{\partial p(\bm s)}{\partial n}\right)dS.
\end{align} 

Let the total pressure $p(\bm r)$ be the sum of the incident and scattered waves at point $\bm r$ on a sufficiently smooth surface of the rigid body. It is expressed by taking the one-sided limit of the sum onto the surface\cite{Bai_2013}. On a sound-hard boundary, where the normal velocity is zero, they are combined to give:
\begin{align}
\frac{1}{2}p(\bm r)
=\int_{\mbox{PAT}} q(\bm s)g(\bm r,\bm s)dS-\int_{\mbox{RB}} p(\bm s)\dfrac{\partial g(\bm r,\bm s)}{\partial n}dS.
\label{eq:helmhuy2}
\end{align}

\subsection*{Discretization of acoustic elements}
The equation (\ref{eq:helmhuy2}) can be discretized by a known protocol used in the boundary element method (BEM)\cite{Sauter2011BoundaryMethods, Bai_2013}. We employ a triangular mesh on the rigid body and apply a continuous piecewise linear function to it, meaning that the dimension of the discretized pressure vector $\bm p$ becomes the number of points on the surface. Each of the phased array elements is represented as a plane so that vector $\bm q$ gives the phased array gain. This discretization gives the following matrix equation in strong form:

\begin{align}
\dfrac12 \bm p&= G\bm q -G_n\bm p,
\end{align}
\noindent where matrices $G$ and $G_n$ represent acoustic connectivity between the surfaces, and correspond to the kernel $g(\bm r,\bm s)$ and $\dfrac{\partial g(\bm r,\bm s)}{\partial\bm n}$, respectively. By employing $B = \dfrac12I+G_n$, it is rewritten as:
\begin{align}
B\bm p &= G\bm q. \label{eq:scat}
\end{align}

The force $F_i$ and the torque $T_i$ along an axis $i \in \{x,y,z\}$ can also be discretized from equations (\ref{eq:force}) and (\ref{eq:torque}) using a normal vector $\bm n$ and a discretized tangential differential operator $D$ on the surface:
\begin{align}
N_i &= \mbox{diag}\left[n_{1, i}, n_{2, i}, \dots \right] \notag\\
M_i &= \mbox{diag}\left[\left(\left(\bm r_1 - \bm c\right) \times \bm n_{1}\right)_i, \left(\left(\bm r_2 - \bm c\right) \times \bm n_{2}\right)_i, \dots\right] \notag \\
F_i &=\bm p^*\left(N_i - \frac{1}{k^2}D_i^*N_iD_i\right)\bm p \label{eq:Df}\\
T_i &=\bm p^*\left(M_i - \frac{1}{k^2}D_i^*M_iD_i\right)\bm p, \label{eq:Dt}
\end{align}
\noindent where $n_l$ denotes normal vector of $l$-th surface element.

\subsection*{Stable levitation condition for rigid body}
Let a small perturbation of position from the target levitation point be $\Delta \bm x=[\Delta x,\Delta y,\Delta z]^T$ and that of rotation be $\Delta \bm \Omega=[\Delta \Omega_x,\Delta \Omega_y,\Delta \Omega_z]^T$.
We combine those two as a generalized position $\bm{\mathcal{X}} =[\Delta x,\Delta y,\Delta z,\Delta \Omega_x,\Delta \Omega_y,\Delta \Omega_z]^T$ and force $\bm{\mathcal{F}} = [F_x,F_y, F_z, T_x, T_y, T_z]^T$.

For sufficiently small $\bm{\mathcal{X}}$, $\bm{\mathcal{F}}$ is expressed in a linear approximation by using a Jacobian matrix $\nabla \bm{\mathcal{F}}$ as: 
\begin{align}
\bm{\mathcal{F}}(\bm{\mathcal{X}})&\simeq \bm{\mathcal{F}(0)}+ \nabla \bm{\mathcal{F}}(\bm{0}) \bm{\mathcal{X}}.
\end{align}
When all the forces and torques acting upon the object balance each other at the levitation point and resistance from the medium is ignored, the equation of motion becomes:
\begin{align}
M\dif[2]{\bm{\mathcal{X}}}{t} = \nabla \bm{\mathcal{F}}(\bm{0}) \bm{\mathcal{X}} \notag\\
\bm{\mathcal{X}}(t) = \bm{\mathcal{X}}(0)e^{\sqrt{M^{-1}\nabla \bm{\mathcal{F}}(\bm{0})}t}, \label{eq:lyapunov}
\end{align}
\noindent where $M=\mathrm{diag}{(mI;\mathcal{I})}$, using $m$ and $\mathcal{I}$ as the mass and inertia matrix of the rigid body, respectively. Here, $\bm{\mathcal{X}}(t)$ is bounded for a bounded initial position $\bm{\mathcal{X}}(0)$ if and only if all eigenvalues of $M^{-1}\nabla\bm{\mathcal{F}(0)}$ are non-positive real numbers. This is called the Lyapunov stability condition in control theory\cite{Strogatz2001NonlinearEngineering}.

\subsection*{Objective function and optimization}
Our goal is to get a phased array drive $\bm q$ that gives the exact force and torque to the body and satisfies the stability condition stated above. This problem is a so-called quadratically constrained quadratic program (QCQP), and many algorithms have been proposed to solve this. However, we empirically employ the following na\"ive unconstrained objective function:
\begin{align}
    \min_{\bm q} \sum_{i=1}^6  w_i  \|\mathcal{F}_i-\tilde{\mathcal{F}}_i\|^2 + v_i \Re[\lambda_i], \label{eq:objfunc}
\end{align}
\noindent where $\bm w, \bm v$ is a positive weight parameter, $\bm{\tilde{\mathcal{F}}}$ is the external force, such as gravity, and torque applied to the rigid body at the levitation point, and $\bm \lambda = \mathrm{eig}(M^{-1}\nabla\bm{\mathcal{F}(0)})$ represents the eigenvalues. Although there is no explicit term to make the imaginary part of the eigenvalues zero in this formulation, minimization of the real part will spontaneously place poles on the real axis.

We employ the L-BFGS method for optimising the algorithm\cite{Liu1989OnOptimization} and use the phase as the explanatory variable, whereas the amplitudes of the transducer are assumed to be constant. The BFGS optimizer has been successfully used before to solve phased array optimization problems\cite{Marzo:2015aa}. 

\subsection*{Numerical expression of the objective function}

Equations (\ref{eq:scat}), (\ref{eq:Df}) and (\ref{eq:Dt}) are combined to give:
\begin{align}
F_i &=\bm q^*G^*B^{-1*}\left(N_i - \frac{1}{k^2}D_i^*N_iD_i\right)B^{-1}G\bm q\\
T_i &=\bm q^*G^*B^{-1*}\left(M_i - \frac{1}{k^2}D_i^*M_iD_i\right)B^{-1}G\bm q.
\end{align}

Also, the coefficients of their Jacobian matrix are expressed as follows:
\begin{align}
\pd{F_y}{x} &= 2\Re\left[\bm q^*G^*B^{-1*}\left(N_y - \frac{1}{k^2}D_y^*N_yD_y\right)B^{-1}\pd{G}{x}\bm q \right]\\
\pd{T_y}{x} &= 2\Re\left[\bm q^*G^*B^{-1*}\left(M_y - \frac{1}{k^2}D_y^*M_yD_y\right)B^{-1}\pd{G}{x}\bm q \right]\\
\pd{F_y}{\Omega_x} &= 2\Re\left[\bm q^*G^*B^{-1*}\left(N_y - \frac{1}{k^2}D_y^*N_yD_y\right)B^{-1}K_x\bm q \right] \notag\\ 
&\quad - \bm q^*G^*B^{-1*}\left(N_z - \frac{1}{k^2}D_y^*N_zD_y\right)B^{-1}G\bm q\\
\pd{T_y}{\Omega_x} &= 2\Re\left[\bm q^*G^*B^{-1*}\left(M_y - \frac{1}{k^2}D_y^*M_yD_y\right)B^{-1}K_x\bm q \right] \notag\\ 
&\quad - \bm q^*G^*B^{-1*}\left(M_z - \frac{1}{k^2}D_y^*M_zD_y\right)B^{-1}\bm q,
\end{align}
\noindent where $\pd{G}{x}$ and $K_x$ are discretized matrices of kernels $\pd{g(\bm r, \bm s)}{x}$ and $(\bm r \times \nabla g(\bm r, \bm s))_x$, respectively. These terms are analytically differentiable with respect to the phase of each transducer. 

\subsection*{Implementation and experimental setup}
We implemented the phased array with commercially available transducers (NIPPON CERAMIC T4010). Each transducer was driven by a 40 kHz rectangular wave of $V_\mathrm{pk-pk}$ = 24 V. 
The rigid bodies were made of extruded polystyrene foam. Each rigid body had a through-hole with a radius of 1 mm, and a pin inserted in the hole was driven by a mechanical arm to move the body to the initial points. The pin was silently removed after the acoustic radiation pressure was applied to the body. 

The optimization parameters were $\bm{w} = (1, 1, 1, 0, 0, 0), \bm v = (1, 1, 1, 0, 0, 0)$ for the sphere with a diameter of 30 mm, and $\bm{w} = (1, 1, 10, 1, 1, 1), \bm v = (1, 1, 1, 1, 1, 1)$ for the regular octahedron with a diagonal length of 50 mm.

Triangular meshes with a mean edge size of 1 mm were employed to ensure sufficient precision in the experiment. A Gauss quadrature rule of order 4 was used for generating coefficients of connectivity matrices. Extended Data Figure 3 shows the resultant force exerted on the sphere by a Gaussian beam. We concluded that this mesh strategy was fine enough, but there is some room to use coarser meshes, which will result in efficient computation costs for each application.

FEM simulations were implemented in the commercially available software COMSOL Multiphysics.
\end{methods}
\begin{addendum}
 \item We thank Dr. S. Hara for suggestions. This work was supported in part by JSPS Grant-in-Aid for Scientific Research (S) 16H06303, JST ACCEL Embodied Media Project (Grant Number JPMJAC1404), and JSPS Grant-in-Aid for JSPS Fellows 15J09604.
 \item[Author Contributions] S.I. designed, developed and implemented the algorithm, devices, experiments and wrote the manuscript; S.M and T.I. implemented the algorithm and conducted the experiments; A.N. developed and implemented the force measurements; Y.M. and H.S. guided the research project; all the authors contributed to the discussion and edited the manuscript.
 \item[Competing Interests] The authors declare that they have no competing financial interests.
 \item[Correspondence] Correspondence and requests for materials should be addressed to Seki Inoue (email: inoue@hapis.k.u-tokyo.ac.jp).
\end{addendum}

\bibliography{mendeley}
\end{document}